\documentclass[%
 reprint,
superscriptaddress,
 amsmath,amssymb,
 aps,
]{revtex4-2}

\usepackage{graphicx}
\usepackage{dcolumn}
\usepackage{bm}

\usepackage{hyperref}
\hypersetup{
    colorlinks=true,
    citecolor=black,
    linkcolor=black,
    urlcolor=black,
}

\begin{document}

\preprint{APS/123-QED}

\title{Counter-streaming heat-flux closure for electron-only collisionless magnetic reconnection}
\author{M. C. McGrae-Menge}
\email[]{madoxmm@physics.ucla.edu} 
\affiliation{Department of Physics and Astronomy, University of California, Los Angeles, CA 90095, USA}
\author{J. R. Pierce}
\affiliation{Department of Physics and Astronomy, University of California, Los Angeles, CA 90095, USA}
\author{M. Almanza}
\affiliation{Department of Physics and Astronomy, University of California, Los Angeles, CA 90095, USA}
\author{A. Velberg}
\affiliation{Plasma Science and Fusion Center, Massachusetts Institute of Technology, Cambridge, MA 02139, USA}
\author{N. Barbour}
\affiliation{Department of Physics, University of Maryland, College Park, Maryland 20742, USA}
\author{W. D. Dorland}
\affiliation{Department of Physics, University of Maryland, College Park, Maryland 20742, USA}
\author{N. F. Loureiro}
\affiliation{Plasma Science and Fusion Center, Massachusetts Institute of Technology, Cambridge, MA 02139, USA}
\author{F. Fiuza}
\affiliation{
GAP/Instituto de Plasmas e Fusao Nuclear, Instituto Superior Tecnico, Universidade de Lisboa, 1049-001 Lisbon, Portugal
}
\author{E. P. Alves}
\email[]{epalves@physics.ucla.edu} 
\affiliation{Department of Physics and Astronomy, University of California, Los Angeles, CA 90095, USA}
\affiliation{Mani L. Bhaumik Institute for Theoretical Physics, University of California at Los Angeles, Los Angeles, CA 90095, USA}

\date{\today}

\begin{abstract}
    In electron-only collisionless magnetic reconnection (MR), a regime of growing importance in turbulent space plasmas, electrons develop strongly non-Maxwellian distributions that invalidate conventional fluid closures based on assumptions of near local thermodynamic equilibrium. Using particle-in-cell (PIC) simulations, we identify the physical origin of the electron heat-flux: counter-streaming between electron sub-populations originating from opposite sides of the current sheet, with each sub-population remaining approximately adiabatic. This insight yields a novel fluid closure, which we implement in fluid simulations using two adiabatic electron fluids initialized on opposite sides of the current sheet. The fluid simulations capture the heat-flux, reconnecting current density, thermal pressure, and bulk flows as observed in PIC, within a reduced fluid description that conventional single-electron-fluid models fundamentally cannot reproduce. The closure is most accurate at low $\beta_{\text{Reconn.}}$ and $B_{\text{Guide}}/B_{\text{Reconn.}}$, regimes relevant to Earth's magnetotail, where it establishes counter-streaming as the physical origin of heat-flux in electron-only collisionless MR and enables its computationally efficient fluid modeling.
\end{abstract}

\maketitle


\section{\label{sec:level1}Introduction}
Magnetic reconnection (MR) is the process by which magnetic field lines topologically rearrange, explosively converting stored magnetic energy into plasma heating, bulk flows, and energetic particles. While MR has been extensively studied in the electron-ion regime \cite{Yamada2010}, recent Magnetospheric Multiscale (MMS) Mission~\cite{Burch2016} observations across diverse turbulent space plasma environments have revealed distinct regions in which the reconnecting sheets are smaller than characteristic ion scales, preventing ion coupling, referred to as ``electron-only MR'' \cite{Phan2018}. Electron-only MR has been identified in Earth's magnetosheath \cite{Phan2018, Stawarz2019,Stawarz2022,Payne2025}, bow shock transition region \cite{Gingell2019,Wang2019,Bessho2022}, and magnetotail \cite{Wang2018,Lu2020,Hubbert2022}. Interestingly, electron-only MR is not simply a ``scaled-down'' version of electron-ion MR, as new features emerge without ions to regulate the dynamics, including dramatically faster reconnection rates \cite{Pyakurel2019, Greess2022, Guan2023, Liu2025, Liu2025PRL} and super-ion-Alfvénic electron jets \cite{Phan2018, Pyakurel2019, Califano2020}. 

Significant progress has been made in understanding kinetic electron behavior in electron-ion MR, providing a foundation from which insights into electron-only MR can be drawn. One recurring and topical kinetic feature is ``electron counter-streaming" -- two coherent electron bulks with different fluid velocities residing at the same spatial location -- which has been observed inside of the electron diffusion region~\cite{Che2009,Fujimoto2009,Bessho2016,Suetrong2026}, plasma outflows/exhausts~\cite{Shuster2014,Hesse2017,Pianpanit2026}, and separatrices~\cite{Drake2003,Cattell2005,Hoshino2021,Shi2022}. Electron counter-streaming drives kinetic instabilities ~\cite{Drake2003,Che2009,Cattell2005}, transport and heating~\cite{Shuster2014,Hoshino2021,Pianpanit2026}, and contributes substantially to the reconnection electric field~\cite{Suetrong2026}. 
Yet current plasma fluid closures -- which truncate the fluid hierarchy and compress kinetic physics into a reduced fluid description helpful for physical insight and computational efficiency -- are unable to capture counter-streaming. The Hammett-Perkins heat-flux closure \cite{Hammett1990}, commonly used to model collisionless Landau damping in MR \cite{Wang2015, Ng2017}, assumes near-Maxwellian distributions, broken by the strongly non-Maxwellian, multi-modal counter-streaming distributions. The Le et al. pressure closure \cite{Le2009,Le2010}, which captures electromagnetic electron trapping in electron-ion MR  \cite{Ohia2012,Egedal2013,Wetherton2019}, is derived for $v_{th}\gg v_{A}$, outside the low $\beta_{\text{Reconn.}}$ regime in electron-only MR where counter-streaming dominates. Counter-streaming's absence from existing fluid closures represents a fundamental limitation in our ability to model electron-only MR with reduced descriptions, which we address in this work.


\begin{figure*}[t]
\centering
\includegraphics[width=\textwidth]{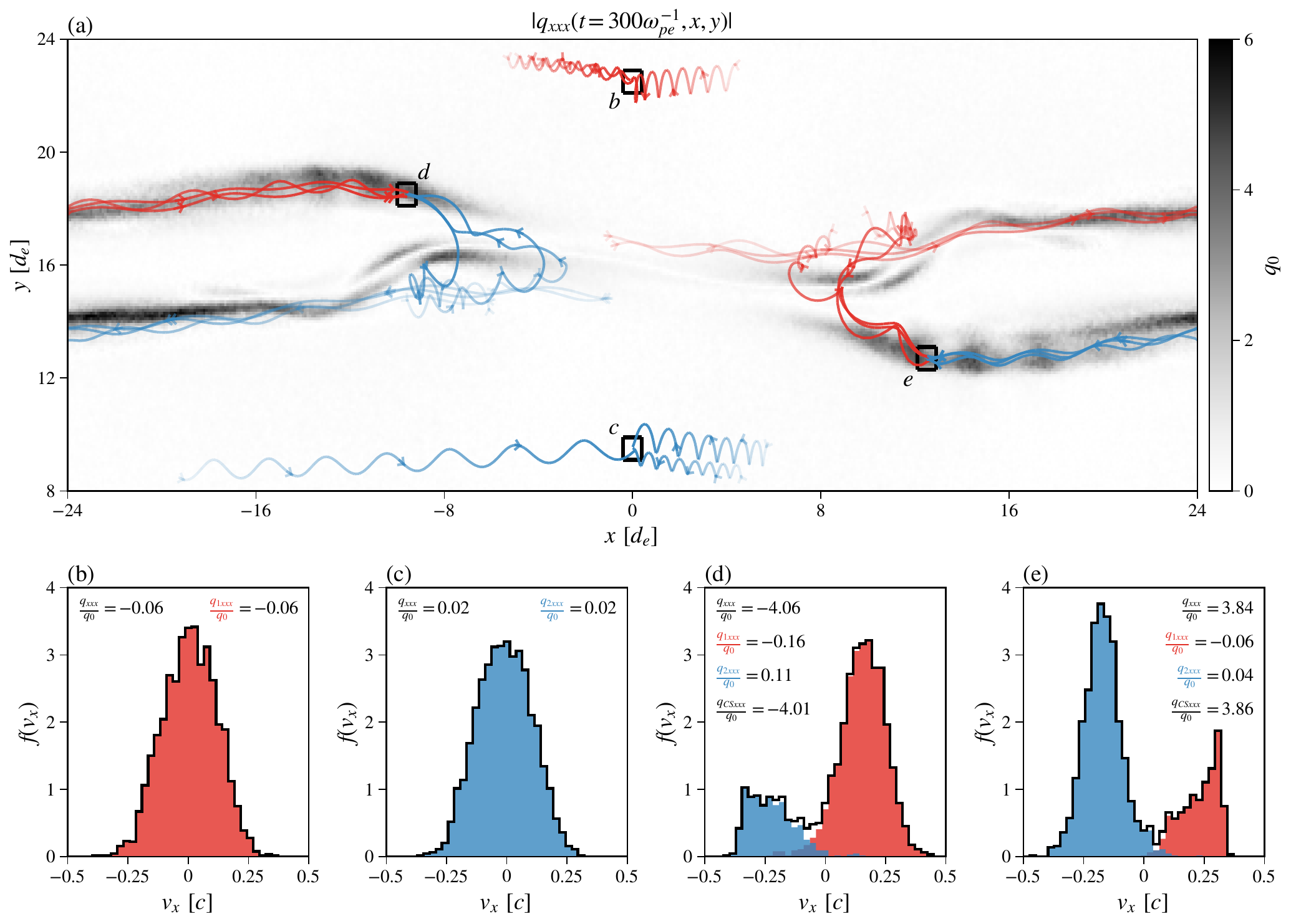}
\caption{\textbf{Heat-flux in electron-only collisionless MR arises from counter-streaming between electron sub-populations originating from opposite sides of the current sheet.} (a) Grayscale map of $|q_{xxx}|/q_0$ at $t=300\omega_{pe}^{-1}$, overlaid with representative electron trajectories from $t=200\omega_{pe}^{-1}$ to $300\omega_{pe}^{-1}$. Trajectories are colored by each particle's $t=0$ position above (red, sub-population 1) or below (blue, sub-population 2) the current sheet mid-plane at $y=16d_e$, with opacity darkening toward the snapshot time. Black boxes b-e mark the locations of the velocity-space analyses shown in (b-e): electron distributions $f(v_x)$ sampled within $0.2d_e\times0.2d_e$ patches at $t=300\omega_{pe}^{-1}$, with the total $f(v_x)$ outlined in black and the sub-population 1 (red) and sub-population 2 (blue) contributions shown separately.  Annotated are local values of $q_{xxx}$, $q_{1xxx}$, $q_{2xxx}$, $q_{CSxxx}$ from Equation \ref{equation4}.}
\label{fig:fig1}
\end{figure*}

In this Letter, using PIC simulations, we show that the electron heat-flux in electron-only collisionless MR arises almost entirely from counter-streaming between populations originating on opposite sides of the current sheet, with each sub-population remaining approximately adiabatic. This insight enables a fluid closure, as the heat-flux is fixed by their relative drift, closing the hierarchy without a third-moment equation. We implement this as a two-electron-fluid closure and show in a forward fluid simulation that it reproduces the PIC heat-flux, current density, pressure, and bulk flows -- which a single-electron-fluid model fundamentally cannot.

We use the fully-kinetic PIC code \textit{OSIRIS} \cite{Fonseca2002,Fonseca2008} to simulate electron-only collisionless MR from first-principles, initializing a 2D-3V force-free double Harris current sheet~\cite{Harris1962} with $\beta_{\text{Reconn.}}=0.1$ and $B_{\text{Guide}}/B_{\text{Reconn.}}=0.1$, representative of magnetotail reconnection conditions where $\beta_{\text{Reconn.}}\sim 0.001-0.1$ and $B_{\text{Guide}}/B_{\text{Reconn.}}\lesssim 0.2$~\cite{Oieroset2023,Richard2025} (Simulation details in Appendix \ref{app:pic_setup}). The electron fluid moments are computed directly from the PIC electron distribution function $f=f(\bm{x},\bm{v},t)$. The density $n(\bm{x},t) = \int f d\bm{v}$, the fluid-velocity $\overline{v}_i(\bm{x},t) = \frac{1}{n}\int v_i fd\bm{v}$, together with the centered thermal pressure and heat-flux tensors respectively:
\begin{equation}\label{equation1}
    p_{ij}(\bm{x},t) = m_e\int (v_i-\overline{v}_i)(v_j-\overline{v}_j)fd\bm{v}
\end{equation}
\begin{equation}\label{equation2}
    q_{ijk}(\bm{x},t) = m_e\int (v_i-\overline{v}_i)(v_j-\overline{v}_j)(v_k-\overline{v}_k)fd\bm{v}
\end{equation}

Figure \ref{fig:fig1} presents a snapshot at $t=300\omega_{pe}^{-1}$, when both $\int |q_{xxx}|d\bm{x}$ and $\max(|q_{xxx}|)$ are near their peak simulation values, $\sim 30\%$ of the magnetic flux has reconnected, and the system has developed well-defined outflows, separatrices, and magnetic islands. Fig. \ref{fig:fig1}a maps $|q_{xxx}|/q_0$; we focus on $\hat{x}$ since it aligns with both the outflow and local $B$ in this low-guide-field regime. Here $n_0$ and $T_0$ are the initial uniform density and temperature, from which we define the characteristic scales $v_{th}\equiv \sqrt{T_0/m_e}$, $p_0\equiv m_en_0 T_0$, $q_0\equiv m_en_0v_{th}^3$, and $J_0\equiv |q_e|n_0c$. $q_{xxx}$ is strongly concentrated in the outflows and separatrices, reaching values up to $\sim 6.6q_0$.

To identify the physical origin of this heat-flux, we examine $f$ in $\{0.2d_e\times0.2d_e\}$ spatial patches at four representative locations marked $b$, $c$, $d$, $e$ in Fig. \ref{fig:fig1}a. In the upstream regions where heat-flux is negligible (Figs. \ref{fig:fig1}b-c), $f$ is approximately Maxwellian, as initialized. In the downstream regions where heat-flux is large (Figs. \ref{fig:fig1}d-e), $f$ is strongly non-Maxwellian, multi-modal, and asymmetric. We now show that a simple physical structure underlies this apparent complexity. 

We tag each electron by its $t=0\omega_{pe}^{-1}$ position above or below the sheet (Fig. \ref{fig:fig1}a, red and blue) and trace representative trajectories to the counter-streaming locations. Electrons from opposite sides of the current sheet cross the mid-plane and converge at the same spatial location with opposing bulk velocities, seen directly in Figs. \ref{fig:fig1}d–e as two well-separated peaks in $f(v_x)$ -- one red, one blue centered at distinct $v_x$. We find the counter-streaming distributions are robust to velocity-space relaxation, with each sub-population retaining its upstream Maxwellian character throughout the reconnection region. We note that above/below electron-fluid decompositions have previously been used to study flow crossover ~\cite{Pianpanit2026} and the pressure supporting the reconnection electric field ~\cite{Suetrong2026}. However, the implications of using such decompositions for understanding the heat-flux or closing the fluid hierarchy have not been previously recognized.

To quantify how this counter-streaming generates the observed heat-flux, we decompose the total distribution into its sub-populations above and below the sheet, $f=f_1+f_2$, focusing on the $\hat{x}$ components (the analysis generalizes straightforwardly to the full tensors  in Appendix \ref{app:full_tensors}). The total density and fluid velocity decompose trivially as $n=n_1+n_2$ and $\overline{v}_x=\frac{n_1\overline{v}_{1x}+n_2\overline{v}_{2x}}{n_1+n_2}$. The higher-order centered moments, however, do not. Defining $\Delta \overline{v}_x \equiv \overline{v}_{1x}-\overline{v}_{2x}$, $\Delta T_{xx} \equiv T_{1xx} - T_{2xx}$ (with $T_{ij}\equiv p_{ij}/n$), and $\Delta n \equiv n_1 - n_2$:
\begin{figure}
\centering
\includegraphics[width=\columnwidth]{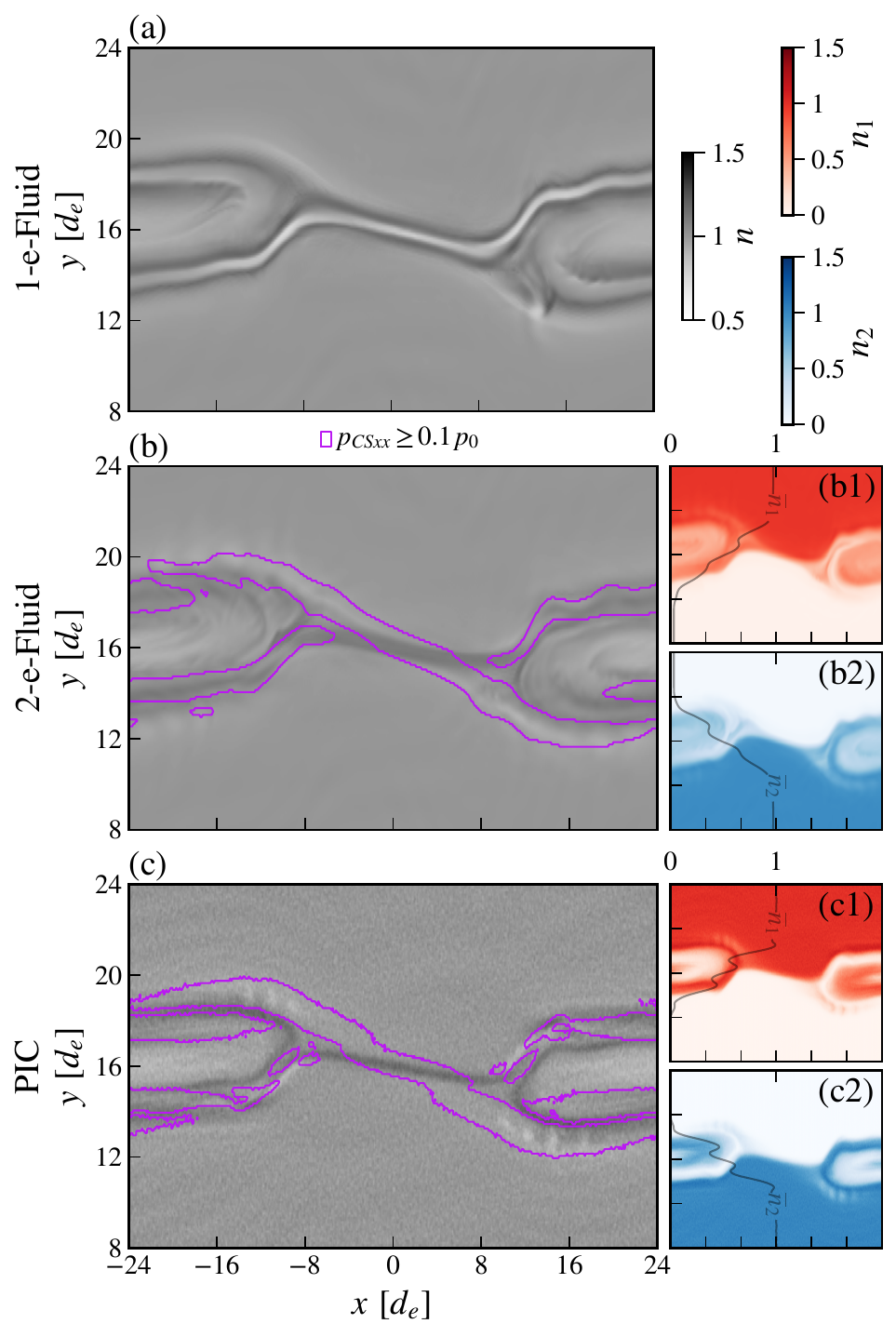}
\caption{\textbf{We construct a 2-e-Fluid model that reproduces the PIC counter-streaming dynamics inaccessible to traditional 1-e-Fluid models.} Electron density $n$ (Grayscale) in 1-e-Fluid (a), 2-e-Fluid (b), and PIC (c) at $t=300\omega_{pe}^{-1}$. Purple lines indicate the boundaries of spatial regions inside of which $p_{CSxx}\geq 0.1p_0$ (Eq. \ref{equation3}), identifying where counter-streaming is present. Side panels (b1, b2) for 2-e-Fluid and (c1, c2) for PIC show the individual sub-population densities $n_1$ and $n_2$, with their $\hat{x}$ averaged profiles $\overline{n}_1$ and $\overline{n}_2$ plotted as a function of $y$. The corresponding initial ($t=0\omega_{pe}^{-1}$) state is shown in Fig.~\ref{fig:fig6} (Appendix~\ref{app:fluid_decomposition}).}
\label{fig:fig2}
\end{figure}
\begin{figure*}[t]
\centering
\includegraphics[width=\textwidth]{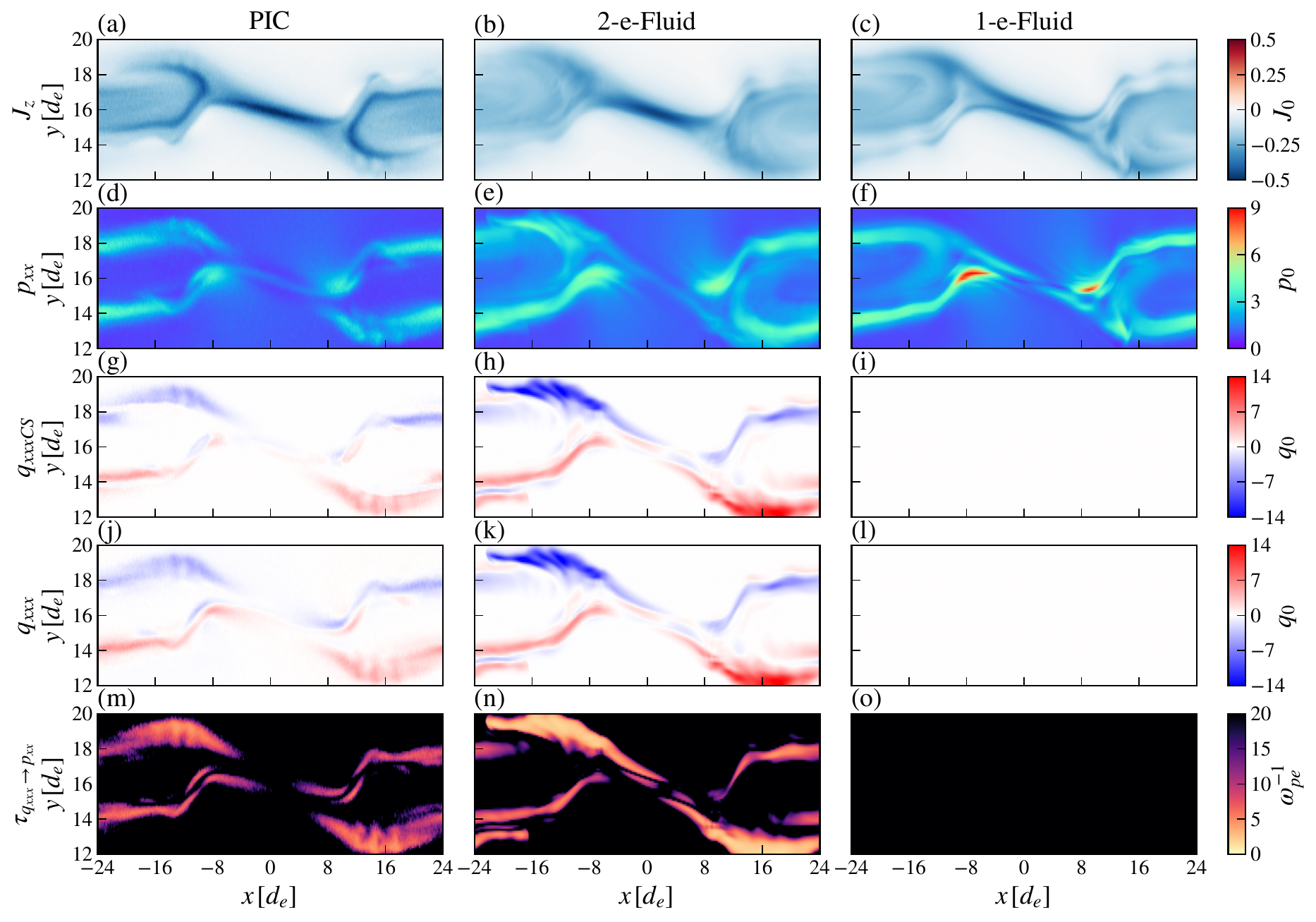}
\caption{\textbf{Our 2-e-Fluid model qualitatively matches PIC across current density, pressure, and heat-flux, which 1-e-Fluid fails to capture.} Snapshots at $t=300\omega_{pe}^{-1}$ in PIC (left-column), 2-e-Fluid (middle), and 1-e-Fluid (right). Rows: reconnecting current density $J_z$ (a-c), thermal pressure $p_{xx}$ (d-f), counter-streaming heat-flux $q_{CSxxx}$ (g-i), total heat-flux $q_{xxx}$ (j-l), and the pressure-modification timescale $\tau_{q_{xxx}\rightarrow p_{xx}}\equiv \frac{p_{xx} d_e}{|q_{xxx}|}$ (m-o). We note that the contribution to $q_{xxx}$ from the isotropization closure is omitted from the fluid panels (k-l), as it is small relative to 2-e-Fluid $q_{CSxxx}$ ($\tau_{Iso} \gg \tau_{q_{xxx}\rightarrow p_{xx}}$), and reconstructing it from the closure would require additional assumptions since only its divergence is constrained.}
\label{fig:fig3}
\end{figure*}

\begin{align}
    p_{xx} &= p_{1xx}+p_{2xx} + \underbrace{m_e\frac{n_1 n_2}{n_1+n_2}\Delta \overline{v}_x^{2}}_{\displaystyle p_{CSxx}} \label{equation3} \\
    q_{xxx} &= q_{1xxx}+q_{2xxx} \nonumber \\
    &\quad + \underbrace{\frac{3 n_1 n_2}{n_1+n_2}\Delta \overline{v}_x\Delta T_{xx} - m_e\frac{n_1 n_2}{(n_1+n_2)^2}\Delta n\Delta \overline{v}_x^{3}}_{\displaystyle q_{CSxxx}} \label{equation4}
\end{align}

\noindent The total thermal pressure and heat-flux (Eqs. \ref{equation3}-\ref{equation4}) acquire additional counter-streaming contributions $p_{CSxx}$ and $q_{CSxxx}$ that cannot be expressed in terms of either sub-population alone. These counter-streaming terms vanish whenever either stream is absent ($n_1=0$ or $n_2=0$) or when the streams co-move ($\Delta \overline{v}_x=0$). Evaluating each term directly from the PIC data in Figs. \ref{fig:fig1}d-e we find $|\frac{q1xxx}{q_{xxx}}|,|\frac{q2xxx}{q_{xxx}}|\lesssim 4\%$, while $q_{CSxxx}$ carries the vast majority of $q_{xxx}$. As we show below, this generalizes across the reconnection region and to all 10 components of $q_{ijk}$. This is the central finding of our work: the sub-populations are individually approximately adiabatic ($q_{1ijk}\approx q_{2ijk}\approx 0$), and the heat-flux is generated almost entirely by their relative motion $q_{ijk}\approx q_{CSijk}$, which closes the fluid hierarchy as $q_{CSijk}$ depends only on lower-order moments of the sub-populations with no third-order contribution. The ``counter-streaming heat-flux closure" follows directly -- partition the electrons above/below the current sheet into two sub-populations and impose adiabaticity on each.

We now test whether a fluid model that explicitly resolves the two sub-populations can reproduce PIC results. We perform 10-moment collisionless, electromagnetic fluid simulations using the code \textit{Gkeyll} \cite{Wang2020}, evolving two electron fluids (2-e-Fluid), one for each sub-population, with the same electron-only MR equilibrium as in PIC. Each fluid carries the 10 moments $n$, $\overline{v}_i$, and the six independent components of the pressure tensor $p_{ij}$. Our closure compresses the electron description by a factor of $\sim156$: from $625$ macroparticles per cell sampling $5D$ phase space in PIC, to just $20$ fluid moments per cell ($10$ per sub-population). Our counter-streaming closure imposes adiabaticity on each sub-population, which we approach numerically through Gkeyll's pressure isotropization scheme $\partial_m q_{ijm}=\kappa v_{th}(p_{ij}-\delta_{ij}p)$ with $p\equiv Tr(p_{ij})/3$, commonly used in electron-ion fluid modeling of collisionless MR ~\cite{Wang2015,Ng2015,Hesse1995,Yin2000}. To approach adiabaticity, we take $\kappa=0.1d_e^{-1}$ -- much smaller than values traditionally used, and as small as possible while retaining numerical stability. This corresponds to an isotropization timescale $\tau_{Iso}\sim\frac{1}{\kappa v_{th}}\approx90\omega_{pe}^{-1}$, much longer than the counter-streaming heat-flux timescale shown later. For comparison, we run a conventional single-electron-fluid (1-e-Fluid) simulation evolving only the total electron moments under the same closure. At $t=0\omega_{pe}^{-1}$, 1-e-Fluid is initialized with uniform density $n_0$, while the 2-e-Fluid model splits this density between the two sub-populations across the current sheet with smooth tanh profiles (Appendix \ref{app:fluid_decomposition}) that approximate the discrete density jump of the PIC decomposition while remaining numerically stable. 



Fig.~\ref{fig:fig2} compares the electron density at $t=300\omega_{pe}^{-1}$ across 1-e-Fluid (Fig.~\ref{fig:fig2}a), 2-e-Fluid (Fig.~\ref{fig:fig2}b), and PIC (Fig.~\ref{fig:fig2}c). The purple curves enclose the regions where $p_{CSxx}\geq 0.1p_0$ (Eq.~\ref{equation3}), marking where counter-streaming is active. In PIC, these regions fill the current sheet, outflows, and separatrices. 2-e-Fluid reproduces this spatial structure in close agreement, while 1-e-Fluid contains no such regions by construction. The side panels show the individual sub-population densities $n_1$ and $n_2$ for 2-e-Fluid (Fig.~\ref{fig:fig2}b1,b2) and PIC (Fig.~\ref{fig:fig2}c1,c2). These side panels show that 2-e-Fluid captures the fraction of each sub-population that crosses to the opposite side, with both the 2D sub-population densities and their $\hat{x}$-averaged profiles $\overline{n}_1$ and $\overline{n}_2$ matching PIC closely. 1-e-Fluid develops a pronounced depletion in the total density at the sheet center, absent in 2-e-Fluid and PIC.

Fig. \ref{fig:fig3} presents a detailed comparison at $t=300\omega_{pe}^{-1}$ across physical quantities that reveal the consequences of missing counter-streaming in the single-fluid model. The reconnecting current density $J_z$ (Fig. \ref{fig:fig3}a-c) shows a striking qualitative failure of 1-e-Fluid: $J_z$ is strongly suppressed at the center of the current-sheet ($y=16d_e$), whereas both 2-e-Fluid and PIC maintain a filled current sheet. This is driven by an underlying density deficit already visible in 1-e-Fluid (Fig. \ref{fig:fig2}a). The origin is geometric: at the mid-plane $\overline{v}_y=0$ by symmetry, blocking 1-e-Fluid transport into the sheet, while the sub-population velocities $\overline{v}_{1y}$ and $\overline{v}_{2y}$ -- present in 2-e-Fluid and PIC but absent by construction in 1-e-Fluid -- carry plasma across from opposite sides supplying density to the layer. The thermal pressure $p_{xx}$ (Figs. \ref{fig:fig3}d-f) exposes a second failure: without a heat-flux channel, 1-e-Fluid develops spurious pressure accumulations in the outflows that are absent in 2-e-Fluid and PIC. In contrast to 1-e-Fluid, the counter-streaming heat-flux $q_{CSxxx}$ (Figs. \ref{fig:fig3}g-i) is large and spatially structured throughout the reconnection region in both PIC and 2-e-Fluid, providing a channel that transports thermal energy away from the outflows and prevents pressure pile-up. 2-e-Fluid reproduces the spatial structure and sign of $q_{CSxxx}$, albeit with a somewhat larger amplitude. The counter-streaming heat-flux $q_{CSxxx}$ in PIC (Fig. \ref{fig:fig3}g) closely resembles the total heat-flux $q_{xxx}$ in PIC (Fig. \ref{fig:fig3}j), confirming that counter-streaming is the dominant heat-flux mechanism across the entire reconnection region -- not only in the localized patches examined in Fig. \ref{fig:fig1}. As shown in Appendix \ref{app:full_tensors} (Fig. \ref{fig:fig5}), this picture also extends to all 10 independent components of $q_{ijk}$, with 2-e-Fluid reproducing the spatial structure and sign of PIC $q_{CSijk}$, which in turn captures the dominant share of the total PIC heat-flux.

We estimate the timescale on which this heat-flux modifies the thermal pressure. From $\partial_t p_{xx}\sim\partial_x q_{xxx}$ with $\partial_x\sim d_e^{-1}$, this timescale is $\tau_{q_{xxx}\rightarrow p_{xx}}\sim p_{xx} d_e/|q_{xxx}|$. Figs. \ref{fig:fig3}m-o show that in PIC and 2-e-Fluid, $\tau_{q_{xxx}\rightarrow p_{xx}}$ reaches down to $\sim 4.1 \omega_{pe}^{-1}$ and $\sim 1.5\omega_{pe}^{-1}$ respectively - comparable to the electron-kinetic timescale on which the system evolves - establishing counter-streaming heat-flux as dynamically important. In 1-e-Fluid the timescale is infinite. 
\begin{figure}[t]
\centering
\includegraphics[width=\columnwidth]{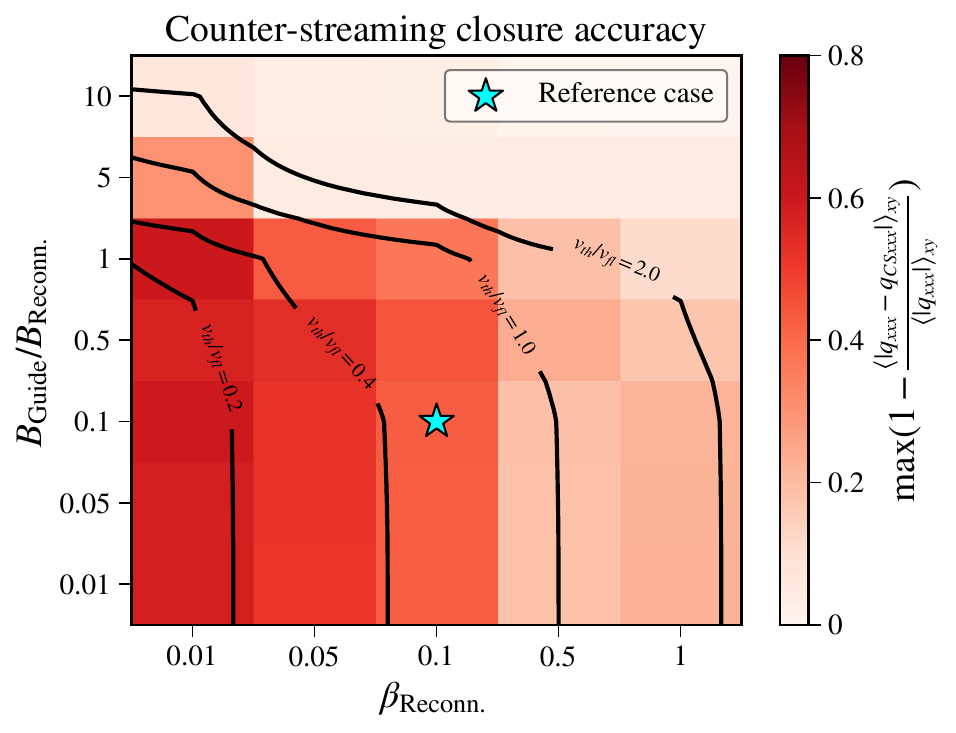}
\caption{\textbf{Our counter-streaming closure is most accurate at low $\beta_{\text{Reconn.}}$ and low $B_{\text{Guide}}/B_{\text{Reconn.}}$, controlled by $v_{th}/v_{fl}$.} Each cell is from an independent PIC simulation at the indicated $\beta_{\text{Reconn.}}$ (varied via $v_{th}$) and $B_{\text{Guide}}/B_{\text{Reconn.}}$ (varied via $B_{\text{Guide}}$). Color gives the time-maximum of $1-\langle|q_{xxx}-q_{CSxxx}|\rangle_{xy}/\langle|q_{xxx}|\rangle_{xy}$, where $\langle\cdot\rangle_{xy}$ denotes the spatial average over the 2D domain; this metric quantifies the fraction of total heat-flux captured by the closure, with darker red indicating better agreement. Black curves are contours of constant initial $v_{th}/v_{fl}$. The blue star marks the reference case $\{\beta_{\text{Reconn.}}=0.1,B_{\text{Guide}}/B_{\text{Reconn.}}=0.1\}$ of Figs. \ref{fig:fig1}-\ref{fig:fig3}.}
\label{fig:fig4}
\end{figure}

To map the regime of validity of the closure, we run a large PIC parameter scan over $\beta_{\text{Reconn.}}$ and $B_{\text{Guide}}/B_{\text{Reconn.}}$, measuring the fraction of total heat-flux captured by the closure (Fig. \ref{fig:fig4}). The closure is most accurate at low $\beta_{\text{Reconn.}}$ and $B_{\text{Guide}}/B_{\text{Reconn.}}$ and degrades smoothly as either increases. The accuracy collapses onto a single parameter, the ratio of the initial electron thermal speed to the maximum $\hat{x}$ initial bulk-flow speed, $v_{th}/v_{fl}$ -- physically natural because the counter-streaming heat-flux scales as $\Delta \overline{v}_x^3$, while intra-population departures from adiabaticity scale as $v_{th}^3$, giving a ratio $(v_{th}/\Delta \overline{v}_x)^3$. Both axes lower this ratio: $\beta_{\text{Reconn.}}$ via $v_{th}$, and smaller $B_{\text{Guide}}/B_{\text{Reconn.}}$ via larger initial drift $v_{fl}$ from the force-free equilibrium.


At later times, a sub-population can circulate back into itself within a magnetic island, developing counter-streaming within a single sub-population that breaks $q_{ijk}\approx q_{CSijk}$ in the 2-e-Fluid model; additional $y$-staggered fluids could help resolve this. The counter-streaming distributions are also unstable to streaming and pressure-anisotropy instabilities driven by $p_{CSij}$, which scatter the sub-populations. Because these instabilities are due to counter-streaming, 2-e-Fluid can capture their linear growth, whereas 1-e-Fluid cannot.

In summary, electron-only collisionless MR develops strongly non-Maxwellian electron distributions that no previously known fluid closure can capture, leaving the regime accessible only to kinetic simulation. We have shown that this complexity is, in fact, the sum of two adiabatic sub-populations originating on opposite sides of the current sheet, whose relative motion sets the electron heat-flux and enables an interpretable and generalizable fluid closure. The resulting computationally efficient 2-e-Fluid simulation reproduces the PIC heat-flux, while 1-e-Fluid develops unphysical density depletion and pressure pile-up. This 2-e-Fluid model can potentially enable efficient fluid modeling of 3D electron-only collisionless MR~\cite{Granier2024}: the $\hat{z}$ heat-flux components are already counter-streaming dominated (Appendix \ref{app:full_tensors}, Fig. \ref{fig:fig5}), and in 3D, where $\partial_z q_{ijz}$ becomes a finite source in the pressure evolution, these components will actively modify the dynamics. This approach could extend to reconnection in turbulence by dynamically identifying forming current sheets ~\cite{Servidio2009,Zhdankin2013,Hu2020,Azizabadi2021} and splitting the fluid across each. More broadly, the above/below fluid initialization and sub-population adiabatic closure presented here can be extended to other reconnection configurations and regimes, such as asymmetric and electron-ion reconnection.

The authors acknowledge the OSIRIS Consortium, consisting of UCLA, University of Michigan, and IST (Portugal) for the use of the OSIRIS 4.0 framework. The authors would also like to acknowledge Liang Wang for helpful discussions regarding the Gkeyll multi-fluid-moment algorithm. This work was supported by the National Science Foundation Grants No. PHY-2108087 and No. PHY-2108089. F.F. acknowledges support from the European Research Council (ERC-2021-CoG Grant XPACE No. 101045172). This research used resources of the National Energy Research Scientific Computing Center (NERSC), a Department of Energy User Facility using NERSC award FES-ERCAPm1157.

\appendix
\section{PIC simulation setup}
\label{app:pic_setup}
We simulate electron-only collisionless magnetic reconnection using the particle-in-cell code \textit{OSIRIS} \cite{Fonseca2002,Fonseca2008}. The 2D-3V domain spans $x\in[-L_x/2,L_x/2]$ and $y\in[0,L_y]$, with $L_x=48d_e$ and $L_y=64d_e$, doubly periodic boundary conditions, resolved at $\Delta x=\Delta y=0.1d_e$. Electron quantities are given in normalized units $|q_e|=c=m_e=n_0=1$, and ions form a static, uniform, neutralizing background ($\frac{m_i}{m_e}\rightarrow \infty$). Each cell is initialized with 625 electron macroparticles using quadratic particle shapes. We choose a timestep $\Delta t=0.06667 \omega_{pe}^{-1}$, which satisfies the Courant condition. All simulations used for Fig. \ref{fig:fig4} are run to $t=1,000\omega_{pe}^{-1} $. 

The equilibrium is a double force-free ($\epsilon_{ijk}J_jB_k=0$) Harris current sheet:

\begin{equation}\label{A1}
    B_x(y)=B_{\text{Reconn.}} \left[   \tanh ( \frac{y-\frac{L_y}{4}}{\lambda}) -
    \tanh( \frac{y-\frac{3L_y}{4}}{\lambda} )-1 \right]
\end{equation}
\begin{equation}\label{A2}
    B_z(y)= \sqrt{B_{\text{Reconn.}}^2+B_{\text{Guide}}^2-B_x^2(y)}
\end{equation}
We use initial current-sheet width $\lambda=1d_e$ and $B_{\text{Reconn.}}=0.5$ ($\Omega_{ce}/\omega_{pe}=0.5$) for all simulations. All of the results shown are from the lower sheet centered at $y=16d_e$. Reconnection is seeded by a small-amplitude perturbation of the form:

\begin{equation}\label{A3}
    B_y(x)=\delta B\{\sin(\frac{2\pi x}{L_x})+\frac{1}{4}\exp{[-(\frac{x-\frac{L_x}{4}}{\frac{L_x}{8}})^2]}\}
\end{equation}

 We use $\delta B=0.01 B_{\text{Reconn.}}$. The Gaussian term in Eq. \ref{A3} deterministically breaks the $x\rightarrow -x$ reflection symmetry of the equilibrium. \textit{Gkeyll} has no mechanism to break this symmetry on its own and would remain exactly $x$-symmetric for all time, while PIC breaks it stochastically via discrete-particle noise; the asymmetric seed forces both codes to select the same reconnection geometry and enables a direct comparison.

The electron fluid velocities are initialized from $\epsilon_{ijk}\partial_j B_k=J_i=q_en_0\overline{v}_i$, i.e. electrons carry the entire current. For the reference case we choose $B_{\text{Guide}}=0.1B_{\text{Reconn.}}$ and $\beta_{\text{Reconn.}}=0.1$, giving $v_{th}=\sqrt{\frac{\beta_{\text{Reconn.}}}{2}}B_{\text{Reconn.}}=0.111803c$. The PIC data and input deck are available on Zenodo~\cite{McGraeMenge2026}.

\begin{figure*}[p]
\centering
\includegraphics[width=\textwidth]{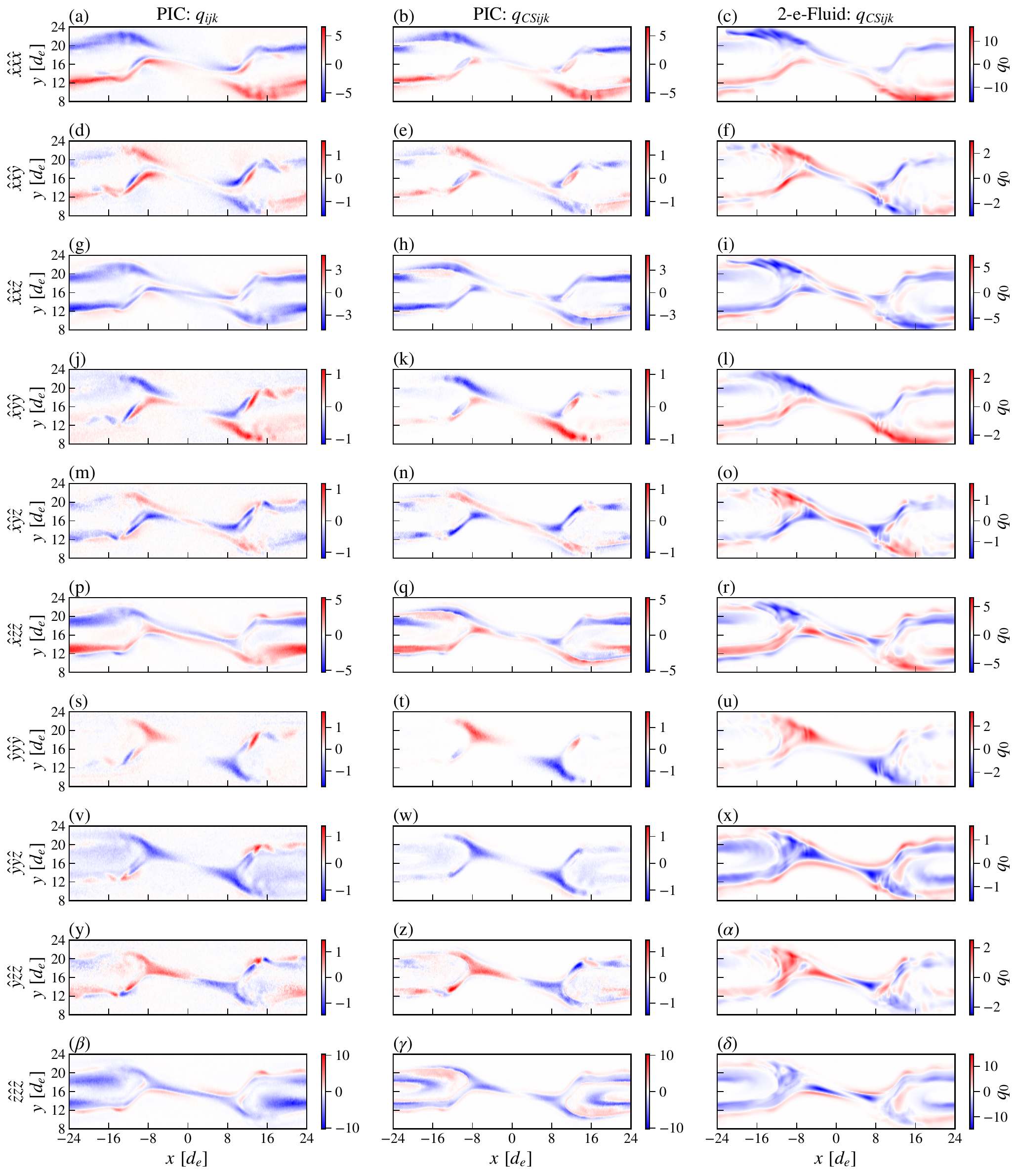}
\caption{\textbf{Counter-streaming is the dominant source of the full PIC heat-flux tensor, and the 2-e-Fluid model reproduces it across all 10 independent components.} Snapshots at $t=300\omega_{pe}^{-1}$, with each row showing one of the 10 independent heat-flux components (labeled $\hat{i}\hat{j}\hat{k}$). The three columns show the total $q_{ijk}$ from PIC (left column), counter-streaming contribution $q_{CSijk}$ from PIC (middle column), and $q_{CSijk}$ predicted by the 2-e-Fluid simulation (right column). The two PIC columns share a common colorbar per row, set to the maximum of $|q_{ijk}|$ from PIC, so amplitudes and structures are directly comparable. The 2-e-Fluid column is on its own per-row colorbar set to the maximum of $|q_{CSijk}|$ from the 2-e-Fluid simulation, to facilitate comparison of spatial structure and sign despite differences in magnitude.}
\label{fig:fig5}
\end{figure*}

\section{Counter-streaming decomposition of the full pressure and heat-flux tensors}
\label{app:full_tensors}
We generalize counter-streaming pressure and heat-flux tensors to full Cartesian coordinates. Using the partition $f=f_1+f_2$, defining $\Delta \overline{v}_i \equiv \overline{v}_{1i}-\overline{v}_{2i}$, $\Delta T_{ij} \equiv T_{1ij} - T_{2ij}$ (with $T_{ij}\equiv p_{ij}/n$), and $\Delta n \equiv n_1 - n_2$:

\begin{equation}\label{equationB1}
    p_{ij} = p_{1ij}+p_{2ij} + \underbrace{m_e\frac{n_1 n_2}{n_1+n_2}\Delta \overline{v}_i\Delta \overline{v}_j}_{\displaystyle p_{CSij}}
\end{equation}
\begin{equation}\label{equationB2}
\begin{aligned}
    q_{ijk} &= q_{1ijk}+q_{2ijk} \\
    &\quad + \underbrace{\frac{n_1 n_2}{n_1+n_2}\Delta \overline{v}_{(i}\Delta T_{jk)} - m_e\frac{n_1 n_2}{(n_1+n_2)^2}\Delta n\,\Delta \overline{v}_i\Delta \overline{v}_j\Delta \overline{v}_k}_{\displaystyle q_{CSijk}}
\end{aligned}
\end{equation}
Where parenthesized indices in Eq. \ref{equationB2} denote $A_{(i}B_{jk)}\equiv A_i B_{jk}+A_j B_{ik}+A_k B_{ij}$. Eqs. \ref{equation3}-\ref{equation4} of the main text are the $ij=xx$ and $ijk=xxx$ components respectively. Figure \ref{fig:fig5} shows all 10 independent components of the total heat-flux $q_{ijk}$ in PIC (left column), the counter-streaming heat-flux $q_{CSijk}$ in PIC (middle column), and the counter-streaming heat-flux $q_{CSijk}$ in 2-e-Fluid (right column) all at $t=300\omega_{pe}^{-1}$. The counter-streaming heat-fluxes are reconstructed from the sub-population moments via Eq. \ref{equationB2}.  We note that 1-e-Fluid predicts $q_{CSijk}=0$ everywhere by construction, so no 1-e-Fluid column is shown. We find that across every component, $q_{CSijk}$ in PIC qualitatively carries the dominant share of the total PIC $q_{ijk}$, establishing counter-streaming as the dominant source of the full heat-flux tensor and a viable model, not just $q_{xxx}$. We find that our 2-e-Fluid model qualitatively reproduces $q_{CSijk}$ (and more importantly $q_{ijk}$), in both spatial structure and sign across all 10 components, with somewhat larger amplitudes that track the same spatial pattern. The weakest agreement between PIC $q_{CSijk}$ and total PIC $q_{ijk}$ is localized to specific regions and components: the current sheet structure for $\hat{x}\hat{x}\hat{z}$ (panels g-h), along the separatrices for $\hat{y}\hat{y}\hat{z}$ (panels v–w), and at the center of the current sheet and inside the plasmoid for $\hat{z}\hat{z}\hat{z}$ (panels $\beta$-$\gamma$). Future work will examine why counter-streaming fails to capture the heat-flux in these regions. 
\section{Fluid simulation setup}
\label{app:fluid_decomposition}

\begin{figure}
\centering
\includegraphics[width=\columnwidth]{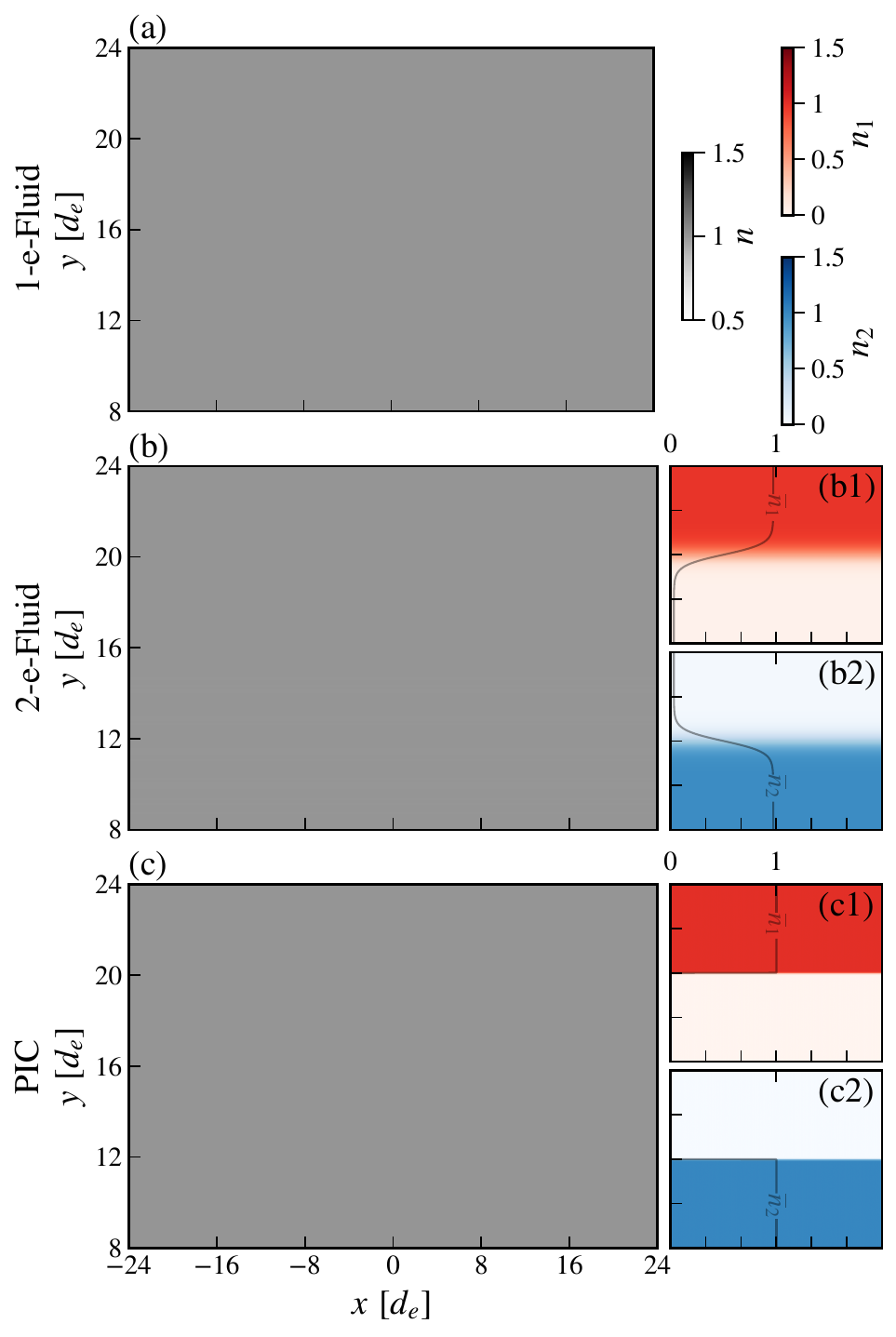}
\caption{\textbf{Initial ($t=0\omega_{pe}^{-1}$) electron density for the 1-e-Fluid, 2-e-Fluid, and PIC simulations.} Total electron density $n$ (grayscale) in 1-e-Fluid (a), 2-e-Fluid (b), and PIC (c). 1-e-Fluid is initialized with uniform density $n_0$, while 2-e-Fluid partitions the density between the two sub-populations across the current sheet using the smooth tanh profiles of Eqs.~\ref{C1},\ref{C2}. Side panels (b1, b2) for 2-e-Fluid and (c1, c2) for PIC show the individual sub-population densities $n_1$ and $n_2$, with their $\hat{x}$ averaged profiles $\overline{n}_1$ and $\overline{n}_2$ plotted as a function of $y$. The PIC sub-populations exhibit a discrete density jump at the sheet, which the 2-e-Fluid initialization approximates with a smooth tanh transition.}
\label{fig:fig6}
\end{figure}
The 1-e-Fluid and 2-e-Fluid simulations of electron-only collisionless MR are performed with the 10-moment electromagnetic, collisionless fluid code \textit{Gkeyll} \cite{Wang2020}, inheriting the PIC domain, grid, boundary conditions, and magnetic-field initial condition (Eqs. \ref{A1}-\ref{A3}) exactly: $x\in[-L_x/2,L_x/2]$ and $y\in[0,L_y]$, with $L_x=48d_e$ and $L_y=64d_e$, doubly-periodic, and $\Delta x=\Delta y=0.1d_e$. As in PIC, ions are a static, uniform, neutralizing background. The fluid equations are advanced with a finite-volume dimensionally-split scheme at CFL fraction $0.66667$. All electron fluids are initialized with $v_{th}=0.111803c$ as in PIC. In 2-e-Fluid, both sub-populations share the same initial fluid velocity $\overline{v}_{1i}=\overline{v}_{2i}=\overline{v}_{i}$ (so that counter-streaming develops self-consistently from the reconnection dynamics rather than being imposed by hand), and also in 1-e-Fluid the single fluid carries $\overline{v}_i$; this common drift is set from $\epsilon_{ijk}\partial_j B_k=J_i=q_en_0\overline{v}_i$. In 1-e-Fluid the electron density is uniform at $n_0$; in 2-e-Fluid the total density is partitioned between the two sub-populations across the current sheet using smooth tanh profiles:
\begin{equation}\label{C1}
    n_1(y) = n_{F} +(\frac{n_0-2n_{F}}{2})\Psi(y)
\end{equation}
\begin{equation}\label{C2}
    n_2(y) = n_0-n_{F} -(\frac{n_0-2n_{F}}{2})\Psi(y)
\end{equation}
\begin{equation}
    \Psi(y)\equiv \tanh(\frac{y-\frac{L_y}{4}}{\lambda_S})-\tanh(\frac{y-\frac{3L_y}{4}}{\lambda_S})
\end{equation}

We choose the smoothing width $\lambda_S=1d_e$ and the density floor $n_F=0.03n_0$; the finite floor and smoothing are needed for numerical stability and we find the resulting dynamics are insensitive to reasonable choices of $\lambda_S$ and $n_F$. These initial density profiles satisfy $n_1(y)+n_2(y)=n_0$ exactly. Fig.~\ref{fig:fig6} shows the resulting $t=0\omega_{pe}^{-1}$ initialization for all three simulations, comparing the uniform 1-e-Fluid density (Fig.~\ref{fig:fig6}a), the smooth 2-e-Fluid partition (Fig.~\ref{fig:fig6}b), and the discrete PIC density split (Fig.~\ref{fig:fig6}c). The fluid data and input decks are available on Zenodo~\cite{McGraeMenge2026}.

One structural limitation of the 2-e-Fluid model is that the free expansion of each sub-population due to its initial pressure gradient into the region occupied by the other is mediated by truncated fluid equations rather than by kinetic effects, and these are not equivalent -- thermal expansion of a plasma is inherently a kinetic process~\cite{Grismayer2008} that fluid dynamics does not properly capture with an adiabatic closure. We note that this difference is subdominant when $\beta_{\text{Reconn.}}\ll 1$, consistent with the regime where our closure is most accurate in Fig. \ref{fig:fig4}, but it does limit the viability of the two sub-population model at large $\beta_{\text{Reconn.}}$.


\bibliography{apssamp}

\end{document}